\documentclass[prd,aps,a4paper,superscriptaddress,twocolumn,nofootinbib]{revtex4}
\usepackage{graphicx}
\usepackage{color}
\usepackage{xcolor}
\usepackage{dcolumn}
\usepackage{bm}
\usepackage{slashed}
\usepackage{amsmath}
\usepackage{latexsym}
\usepackage{amssymb}
\usepackage{mathrsfs}
\usepackage{amsfonts}
\usepackage{xspace}

\allowdisplaybreaks

\newcommand{\iu}{\mathrm{i}}   

\begin{document}
\title{Null test of general relativity based on gravitational waves}

\author{Dawen Chen}
\email[Dawen Chen:~]{chendawen23@mails.ucas.ac.cn}
\affiliation{School of Fundamental Physics and Mathematical Sciences, Hangzhou Institute for Advanced Study, University of Chinese Academy of Sciences, 
Hangzhou 310024, China}
\affiliation{Institute of Theoretical Physics, Chinese Academy of Sciences, Beijing 100190, China}
\affiliation{University of Chinese Academy of Sciences, Beijing 100049, China}
\author{Zeyuan Kang}
\email[Zeyuan Kang:~]{202321160004@mail.bnu.edu.cn}
\affiliation{School of Physics and Astronomy, Beijing Normal University, Beijing 100875, China}
\author{Zhoujian Cao\footnote{corresponding author}}
\email[Zhoujian Cao:~]{zjcao@bnu.edu.cn}
\affiliation{School of Fundamental Physics and Mathematical Sciences, Hangzhou Institute for Advanced Study, University of Chinese Academy of Sciences, 
Hangzhou 310024, China}
\affiliation{School of Physics and Astronomy, Beijing Normal University, Beijing 100875, China}

\begin{abstract}
Several hundreds of gravitational wave events have been detected. They are good tools to investigate various fundamental problems, including tests of general relativity (GR) theory. Among the GR test schemes based on gravitational wave data, the parameterized scheme has been extensively studied by many groups including the LIGO-VIRGO-KAGRA collaboration (LVK). Existing results indicate that the deviation parameters from GR are small. In the current paper, we would like to estimate whether these deviation parameters are small enough to confirm GR. We transform the deviation parameters into mass--mass relations and, by comparing the resulting mass--mass relations from different parameters, we can estimate the probabilities that the data favor the GR theory. Applying our null test scheme to five O1/O2 binary black hole events, we find some GR deviation signatures but they are not conclusive yet. We also find no monotonic correlation between the network SNR and the strength of GR deviation, suggesting that the observed tensions are not simply driven by signal strength. Our null test scheme can be easily generalized and applied to all gravitational wave observation data.
\end{abstract}

\maketitle

\section{Introduction}
Together with the classical tests of GR \cite{Will2006,2018tegp.book.....W,Bertotti2003,Fienga2024} including the light bending, the gravitational redshift, perihelion precession of binary orbits and frame dragging of rotating bodies, modern tests including cosmology, gravitational waves and black hole shadows have shown highly consistent results with GR so far. Although GR remains the most well-established gravitational theory, there are many implications indicating that GR is not the final theory describing gravity. Looking for the validity boundary of general relativity (GR) theory is one of the most important problems in fundamental physics. There are two categories of methods to investigate such task. One is constructing a theory beyond GR and comparing the predictions of that theory and GR with observations. This approach is analogous to the historical development of GR from Newtonian gravity. The other category method is conducting a null test whose null hypothesis is established by GR. A similar \textit{approach} was used to rule out the \textit{ether}, which inspired the emerging special relativity theory. Broadly speaking, the first category is more model-dependent than the second.

As ground-based gravitational wave (GW) detectors including LIGO, VIRGO and KAGRA (LVK) have completed four observation runs, gravitational wave astronomy has emerged as a cutting-edge field which provides unique information with a brand-new messenger. Gravitational waves have been used as an unparalleled tool \cite{PhysRevD.80.122003,PhysRevD.85.082003,PhysRevD.89.082001,PhysRevLett.116.221101,PhysRevX.6.041015,PhysRevLett.123.011102,PhysRevD.100.104036,PhysRevD.107.044020} to test existing gravitational theories and probe new physics \cite{2025PhRvD.112h4080A,Arun2022}.

According to the standard data analysis scheme of gravitational waves, matched filtering, waveform models are required to perform the GR test based on gravitational wave detections. Some waveform models depend on specific gravity theories such as Brans-Dicke theory, Lorentz violation theory, parity violation theory and others \cite{PhysRevD.87.104029,2020PhRvD.102l4035L,universe7070235,PhysRevD.110.064044,PhysRevD.111.104012,AMARILO2024138785,WANG2022137416,An2023,SUN2024138350,AN2025102062,PhysRevD.109.104062,2022ApJ...930..139Z}.
Some other models extend the GR waveform phenomenologically without relying on any specific gravity theories. These waveform models include ppE framework \cite{PhysRevD.80.122003}, the Test Infrastructure for General Relativity (TIGER) framework \cite{PhysRevD.85.082003,PhysRevD.89.082001}, the Flexible Theory-Independent (FTI) framework \cite{PhysRevD.107.044020}, parametrized quasi-normal mode framework \cite{2024PhRvD.110j4007C} and others.

In order to strengthen the model-independent test of GR based on gravitational wave detection, people designed a consistency test scheme based on GR waveforms. If the detected signal is completely consistent with GR prediction, one expects the following features. The residual signal after subtracting the GR-predicted waveform should be zero \cite{PhysRevLett.116.221101,PhysRevD.112.084080}. Source parameters estimated through different stages of signals should be consistent and the resulting parameters difference should be zero \cite{PhysRevLett.116.221101,PhysRevD.112.084080}. The extra parameters involved in the waveform formula beyond the predictions by GR should be zero \cite{PhysRevLett.116.221101,PhysRevX.6.041015,PhysRevLett.123.011102,PhysRevD.112.084080,PhysRevD.110.124062,59zd-qvbd}. The number of extra dimensions of spacetime should be zero \cite{PhysRevLett.123.011102}. The resulted luminosity distance should be the same as measured by electromagnetic waves \cite{PhysRevLett.123.011102,An2023}. The extra polarization modes of gravitational waves should not exist \cite{PhysRevLett.123.011102}. There should be no modified dispersion relation to the GR prediction \cite{PhysRevLett.118.221101}. These vanishing behaviors provide a null test of GR based on GR waveform models.

The authors of \cite{PhysRevD.74.024006,2006CQGra..23L..37A,PhysRevD.82.064010} proposed a little bit different idea to the above mentioned consistency test of GR. If the detected signal is completely consistent with GR prediction, some coefficients involved in the waveform should be completely determined by the source parameters. Moreover, the determined values of these coefficients should be the values as GR predicts. Even if some systematic biases introduced by the detection instrument contaminate the detected source parameters, the aforementioned relation between the waveform coefficients and the GR-predicted values is still valid. Such consistency can be used for the null test of GR. The idea is similar to the mass--mass diagram used by pulsar binary observations to test GR \cite{2015IJMPD..2430018M,2023GReGr..55...76L,2024LRR....27....5F}.

The GR testing idea of \cite{PhysRevD.74.024006,2006CQGra..23L..37A,PhysRevD.82.064010} has not been applied to real detection data yet. And more the existing work is valid only for the inspiral stage. In the current paper, we firstly extend this interesting idea to full inspiral-merger-ringdown signals and then use the Binary Black Hole (BBH) events in O1 and O2 runs of LIGO-VIRGO-KAGRA collaboration to test GR.

The arrangement of the rest of the current paper is as follows. We will explain the waveform models and the corresponding null test scheme in the next section.
We verify the methodology of the null test with the ten O1/O2 BBH events' data (using IMRPhenomXAS, $\psi_{j(l)}$), then we apply the null test scheme to LVK's posterior data of five O1/O2 BBH events listed in Sec.~\ref{sec3} (using IMRPhenomPv2, $\phi_{j(l)}$).
The details of the null test scheme are explained there. Finally we conclude the paper in the last section and some discussion is also presented there.

Throughout the paper we use geometrized units where $G=c=1$.

\section{Modified waveform models and null test scheme of GR}\label{sec2}
The dynamical evolution of compact object binary coalescence (CBC) systems emitting GWs can be roughly partitioned into different stages in the frequency domain, Inspiral-Intermediate-Merger-Ringdown. Various approaches have been devised to model GW waveforms, thereby generating several families of waveform models \cite{2023arXiv231101300L}. To address the requirement of computational efficiency, researchers have designed phenomenological model families to speed up GW data analyses including parameter estimations (PE). Two main phenomenological families, SEOBNR and IMRPhenom, play a crucial role in data processing and PE pipelines. Formally, each waveform model is characterized by a set of post-Newtonian (PN) coefficients. We would like to emphasize that these PN coefficients have been calibrated to numerical relativity results and they are valid for the full inspiral-merger-ringdown process.

Since each waveform model used by LVK calibration is similar for our null test scheme of GR, we just use the IMRPhenomXAS waveform model as an example to illustrate our method. As in the original IMRPhenomXAS waveform model, we only consider the (2,-2) spherical harmonic mode \cite{PhysRevD.78.124011} in the modified IMRPhenomXAS waveform model
\begin{align}
\tilde{h}_{2-2}(f,\bm{\theta}) = A(f,\bm{\theta}) e^{-\iu\phi(f,\bm{\theta})},f > 0\label{eq1}
\end{align}
where $f$ is the gravitational wave frequency, $\bm{\theta}$ are the waveform parameters including the total mass $M$, the symmetric mass ratio $\eta$ and black holes' spins $\chi_{1,2L}$, $A(f,\bm{\theta})$ corresponds to the waveform amplitude and $\phi(f,\bm{\theta})$ corresponds to the waveform phase. The plus and cross polarizations $H_+$ and $H_\times$ can then be obtained from the modelled $\tilde{h}_{2-2}$ according to the relation between $\tilde{h}_{22}$ and $\tilde{h}_{2-2}$ \cite{PhysRevD.77.024027,PhysRevD.102.064001}
\begin{align}
        \label{equations:h2-2_to_hpc}
        H_+ &= + \sqrt{\frac{5}{16\pi}} \frac{1+\cos^2\vartheta}{2} \cdot \tilde{h}_{2-2}e^{-2\iu\varphi} \\
        H_\times &= -\iu \sqrt{\frac{5}{16\pi}} \cos\vartheta \cdot \tilde{h}_{2-2}e^{-2\iu\varphi}
\end{align}
where $\vartheta$ and $\varphi$ are the inclination angle and the azimuthal angle of the BBH respectively.

\subsection{Brief review of the IMRPhenomXAS waveform model}
The amplitude and the phase defined in (\ref{eq1}) are modeled in different ansatzes during different stages \textbf{Ins}piral-\textbf{Int}mediate-\textbf{M}erger-\textbf{R}ingdown as in IMRPhenomXAS
\begin{align}
        A(f,\bm{\theta}) &=
        \begin{cases}
            A_\text{Ins}(f_\text{Ins}^A,\bm{\theta}), f_\text{Ins}^A \in (0, f_T^A]\\
            A_\text{Int}(f_\text{Int}^A,\bm{\theta}), f_\text{Int}^A \in (f_T^A, f_\text{peak}]\\
            A_\text{MR}(f_\text{MR}^A,\bm{\theta}) , f_\text{MR}^A \in (f_\text{peak}, 0.3/M)
        \end{cases},\nonumber\\
        \phi(f,\bm{\theta}) &=
        \begin{cases}
            \varphi_\text{Ins}(f_\text{Ins}^\varphi,\bm{\theta}), f_\text{Ins}^\varphi \in (0, f_\text{MECO} - \delta_R]\\
            \varphi_\text{Int}(f_\text{Int}^\varphi,\bm{\theta}), f_\text{Int}^\varphi \in (f_\text{MECO} - \delta_R, f_T^A + \delta_R]\\
            \varphi_\text{MR}(f_\text{MR}^\varphi,\bm{\theta}) , f_\text{MR}^\varphi \in (f_T^A + \delta_R, 0.3/M)
        \end{cases}.\nonumber\\
f_T^A&\equiv f_\text{MECO}+\frac{1}{4}(f_\text{ISCO}-f_\text{MECO}),\\
\delta_R&\equiv0.03(f_T^\varphi-f_\text{MECO}),\\
f_T^\varphi &\equiv 0.6(\frac{1}{2}f_\text{RD} + f_\text{ISCO}),\\
f_\text{peak}&\equiv\left|f_{\rm RD}+f_{\rm damp}\sigma\frac{\sqrt{1-\lambda}-1}{\lambda}\right|.
\end{align}
In the original IMRPhenomXAS waveform model, MECO means minimal energy circular orbit \cite{PhysRevD.95.064016}, ISCO means innermost stable circular orbit, ``peak'' means the peak of the ringdown \cite{Khan:2015jqa} and $f_{\rm RD}+\iu f_{\rm damp}$ corresponds to the quasinormal modes of the final remnant black hole. Here $0.3/M$, corresponding to the condensed notation $0.3Mf$ in the original paper \cite{PhysRevD.102.064001}, denotes the cutoff frequency implemented for IMRPhenomXAS in the LSC Algorithm Library Suite (LALSuite) \cite{lalsuite,Iacovelli_WF4Py_2022}. Notations $f_\text{MECO}$, $f_\text{ISCO}$, $f_\text{RD}$, $f_\text{damp}$, $\sigma$ and $\lambda$ are all functions of the symmetric mass ratio $\eta$ and black holes' spins $\chi_{1,2L}$. The expressions of these functions are all determined by phenomenological numerical fittings. In the current paper, since we discuss the situations beyond GR, these variables no longer hold such physical meanings. Instead we just follow the IMRPhenomXAS waveform model and take the same values for these variables.

The phase expressions listed in the above formulae are
\begin{align}
&\varphi_\text{Ins} = \varphi_\text{TF2}(Mf;\bm{\theta}) + \frac{1}{\eta}\left(
 \sigma_0 + \sigma_1f + \frac{3}{4}\sigma_2f^{4/3} \right.\nonumber\\
&\left.+ \frac{3}{5}\sigma_3f^{5/3}
 + \frac{1}{2}\sigma_4f^{6/3} + \frac{3}{7}\sigma_5f^{7/3}
\right), \label{equation:IMRPhenomXAS_ins_phase_ansatz}\\
&\varphi_\text{TF2}(Mf;\bm{\theta}) =
            2\pi ft_c - \varphi_c - \frac{\pi}{4} \nonumber\\
&+ {\frac{3}{128\eta}} (\pi Mf)^{-5/3} \sum_{i=0}^{i=7}\varphi_i(\bm{\theta})(\pi Mf)^{i/3},\label{eq2}\\
&\eta \cdot \varphi_\text{Int}^\prime = b_0 + b_4f^{-4} + b_3f^{-3} + b_2f^{-2} + b_1f^{-1}\nonumber\\
                                            &- \frac{4c_0a_\varphi}{(f - f_\text{RD})^2 + (2f_\text{damp})^2},\\
&\eta \cdot \varphi_\text{MR}^\prime = c_\text{RD} + \sum_i^n c_if^{-p_i} + \frac{c_0a_\varphi}{f_\text{damp}^2 + (f - f_\text{RD})^2},
\end{align}
where $\varphi_\text{Int}$ and $\varphi_\text{MR}$ are determined by the derivatives $\varphi_\text{Int}^\prime$ and $\varphi_\text{MR}^\prime$ with respect to the gravitational wave frequency $f$. Here $\sigma_i$, $b_i$ and $c_i$ are all functions of the waveform parameters $\bm{\theta}$. Their expressions are also determined by phenomenological numerical fittings in IMRPhenomXAS. We adopt such expressions directly.

To ensure the phenomenological models are physically reasonable, phenomenological parameters appearing in the phenomenological amplitude $A(f,\bm{\theta})$ and phase $\varphi(f,\bm{\theta})$ are required to be $C^0$-continuous and $C^1$-continuous respectively at the frequency point where inspiral and intermediate, or intermediate and merger-ringdown phases are glued together. The modified inspiral phase is concatenated with the subsequent phase continuously at the frequency point $f_\text{MatchIN} \equiv f_\text{MECO} - \delta_R$. Specifically, the $C^1$ continuity is required, meaning the values of $\varphi_\text{Ins}$ and $\varphi_\text{Int}$ and their derivatives $\varphi^\prime_\text{Ins}$ and $\varphi^\prime_\text{Int}$ must be equal at $f_\text{MatchIN}$ respectively. One pair of additional integral coefficients, $C_{1,\text{Int}}$ and $C_{2,\text{Int}}$, are introduced to eliminate the inherent mismatches which come from separate numerical fittings of the inspiral phase and the intermediate phase
\begin{align}
C_{2,\text{Int}} &\equiv \varphi^\prime_\text{Ins}(f_\text{MatchIN}) - \varphi^\prime_\text{Int}(f_\text{MatchIN}), \nonumber\\
C_{1,\text{Int}} &= \varphi_\text{Ins}(f_\text{MatchIN}) - \varphi_\text{Int}(f_\text{MatchIN}) - C_{2,\text{Int}} f_\text{MatchIN}.\nonumber
\end{align}

\subsection{Parameterized IMRPhenomXAS waveform model}
In GR the coefficients $\varphi_i(\bm{\theta})$ of the polynomial structure in $(\pi Mf)^{1/3}$ appearing in the CBC inspiral (\ref{eq2}) are the post-Newtonian (PN) coefficients.
Their mathematical expressions are derived purely analytically from the post-Newtonian theory of general relativity, therefore encoding the predictions and the theoretical structures of GR---a unique opportunity and scenario to utilize existing GW observation data to test GR. Moreover, specific physical phenomena (such as dipole radiations and massive gravitons), or particular properties of other gravity theories beyond GR (such as Chern-Simons modified gravity), which are expected to be encoded in the theoretical GW waveforms, will ultimately manifest as deviations in several PN coefficients \cite{PhysRevD.80.122003}. PN coefficients thus are granted special significance.

In order to extend the IMRPhenomXAS waveform model to parameterized IMRPhenomXAS waveform model to describe gravity theories beyond GR, we replace the PN coefficients involved terms as follows
\begin{align}
        &\quad \frac{3}{128\eta} (\pi Mf)^{-5/3} \sum_{i=0}^{i=7}\varphi_i(\bm{\theta})(\pi Mf)^{i/3} \notag\\
        &\rightarrow \sum_{j=0}^{j=7} \left[ \psi_j(\bm{\theta}) + \psi_{jl}(\bm{\theta}) \cdot \ln{f} \right] \cdot f^{(j-5)/3},\label{eq5}
\end{align}
where each $\psi_{j(l)}$ can be further decomposed into the ``spin-independent'' $\alpha_{j(l),\text{NS}}(\bm{\theta})$ and the ``spin-dependent'' $\alpha_{j(l),\text{S}}(\bm{\theta})$ with a common factor $\text{Coeff}(M,\eta)$
\begin{align}
\psi_j &= \text{Coeff}(M,\eta) \cdot (\alpha_j + \alpha_{jl}\cdot\ln{(\pi M)}), \\
\psi_{jl} &= \text{Coeff}(M,\eta) \cdot \alpha_{jl},\\
\text{Coeff}(M,\eta) &\equiv \frac{3}{128}\frac{1}{\eta} \cdot (\pi M)^{(j-5)/3}, \\
\alpha_j &= \alpha_{j,\text{NS}} + \alpha_{j,\text{S}}, \label{eq3}\\
\alpha_{jl} &= \alpha_{jl,\text{NS}} + \alpha_{jl,\text{S}}. \label{eq4}
\end{align}
These notations focus on the Fourier frequency dependence $f^{1/3}$ instead of $(\pi Mf)^{1/3}$ in GR case. Also notice that $\varphi_{j(l)}$ of some orders contains an extra dependence on $\ln{f}$ except the $f^{1/3}$ power-law dependence that we will regard them as two kinds of PN coefficients and distinguish between them with an ``$l$'' subscription (integer-order-type $\psi_j$ and log-order-type $\psi_{jl}$). For convenience we list the explicit expressions of each ``$\alpha$'' in the Appendix~\ref{appendix:pn_coefficient_analytic_expressions}.

Ideally, all deviations PN coefficients should be allowed to vary so that all possible non-GR deviations can be included at a time, like a fully parameterized ppE waveform model. However, due to the high degrees of freedom in parameter estimation and especially the mutual compensation among different PN coefficients within the same inspiral phase polynomial structure, the resulting inferences tend to be overly conservative and inaccurate. To obtain relatively more meaningful results, the single-parameter test is expected to be a more effective strategy. Accordingly, we choose to deviate a single coefficient at a time to form a parameterized IMRPhenomXAS waveform model.

More specifically, we pick out one of the parameters $\psi_j$ and $\psi_{jl}$ appearing in (\ref{eq5}) and let it vary freely together with other GR waveform parameters including the total mass, the symmetric mass ratio and the black holes' spins.

Compared with GR waveform, the non-GR parameter $\psi_{j(l)}$ results in a phase correction $\Delta \varphi_{j(l),\text{IN}}(f)$ for the inspiral phase. Respectively for $\psi_j$ and $\psi_{jl}$, we have
\begin{align}
&\Delta \varphi_{j,\text{IN}}(f) = (\psi_{j}-\psi_{j}^{\rm GR}(M,\eta;\chi_{1L},\chi_{2L})) f^{(j - 5)/3},\\
&\Delta \varphi_{jl,\text{IN}}(f) = \left(\psi_{jl}-\psi_{jl}^{\rm GR}(M,\eta;\chi_{1L},\chi_{2L})\right) f^{(j - 5)/3}\ln f.
\end{align}
The resulting derivatives read as
\begin{align}
&\Delta \varphi^\prime_{j,\text{IN}}(f) = \frac{j-5}{3}(\psi_{j}-\psi_{j}^{\rm GR}(M,\eta;\chi_{1L},\chi_{2L})) f^{(j - 8)/3},\\
&\Delta \varphi^\prime_{jl,\text{IN}}(f) = \left(\psi_{jl}-\psi_{jl}^{\rm GR}(M,\eta;\chi_{1L},\chi_{2L})\right) \times\nonumber\\
&\left(\frac{j-5}{3}f^{(j - 8)/3} \ln{f} + f^{(j - 8)/3}\right).
\end{align}
Correspondingly the two integral constants $C_{1,\text{Int}}$ and $C_{2,\text{Int}}$ are also corrected by
\begin{align}
\Delta C_{2,\text{Int}} &= \Delta \varphi^\prime_{j(l),\text{IN}}(f_\text{MatchIN}), \\
\Delta C_{1,\text{Int}} &= \Delta \varphi_{j(l),\text{IN}}(f_\text{MatchIN})-\Delta C_{2,\text{Int}}f_\text{MatchIN}.
\end{align}

In order to preserve the continuity of the waveform, the above $\Delta \varphi_{j(l),\text{IN}}(f)$ is added to the inspiral phasing and the additional accumulative phasing difference resulting from $\Delta C_{1,2,\text{Int}}$ is added to the total phasing of intermediate-ringdown. Consequently we have the final phase behavior for our parameterized IMRPhenomXAS waveform model as
\begin{align}
\varphi_\text{Ins}^{j(l)}(f) &= \varphi_\text{Ins}^{\text{GR}}(f) + \Delta \varphi_{j(l),\text{IN}}(f),\\
\varphi_\text{Int}^{j(l)}(f) &= \varphi_\text{Int}^{\text{GR}}(f) + \Delta C_{2,\text{Int}}f+\Delta C_{1,\text{Int}},\\
\varphi_\text{RD}^{j(l)}(f) &= \varphi_\text{RD}^{\text{GR}}(f) + \Delta C_{2,\text{Int}}f+\Delta C_{1,\text{Int}}.
\end{align}

Since the non-GR parameters $\psi_{j(l)}$ do not affect the amplitude part, we keep the amplitude behavior the same as GR waveform. Instead of $\psi_{j(l)}$, it is also convenient to use the fractional deviations $\delta_{j(l)}$ to denote the parameter deviation fraction
\begin{align}
\delta_{j(l)}\equiv\frac{\psi_{j(l)}-\psi_{j(l)}^{\rm GR}}{\psi_{j(l)}^{\rm GR}}.
\end{align}
When $\delta_{j(l)}=0$ the GR waveforms are recovered.

\subsection{Null test scheme of GR based on the parameterized IMRPhenomXAS waveform model}\label{secIIC}

Regarding GR, if we have knowledge of the value of a given $\psi_{j(l)}$, the functional form of this $\psi_{j(l)}$ with respect to the BBH parameters will relate $m_2$ to $m_1$ and black holes' spins. In addition, we note that $\psi_{0}$, $\psi_{1}$, $\psi_{2}$, $\psi_{0l}$, $\psi_{1l}$, $\psi_{2l}$, $\psi_{3l}$, $\psi_{4l}$, $\psi_{6l}$ and $\psi_{7l}$ do not depend on black holes' spins.

Since the PN coefficients $\psi_1=\psi_{0l}=\psi_{1l}=\psi_{2l}=\psi_{3l}=\psi_{4l}=\psi_{7l}=0$ in GR theory, these coefficients can not be used to relate $m_2$ to $m_1$ and black holes' spins.

We would like to borrow the GR null test idea through the mass--mass diagram used in pulsar timing, and $\psi_0$, $\psi_2$ and $\psi_{6l}$ are chosen for IMRPhenomXAS for their spin-independence and nonconstancy. The numerical analysis in Sec.~\ref{sec3} uses the dimensionless PN coefficients $\phi_{j(l),\text{NS}}$ for the LVK collaboration's IMRPhenomPv2 results; the corresponding spin-independent and nonconstant coefficients there are $\phi_{2,\text{NS}}$, $\phi_{4,\text{NS}}$, $\phi_{5l,\text{NS}}$, $\phi_{6,\text{NS}}$, and $\phi_{7,\text{NS}}$. The notation convention and the transfer between the two parametrizations are discussed in detail in Sec.~\ref{sec3}.

The detailed inverse-mapping procedure for constructing the equi-coefficient curves has been described by \cite{PhysRevD.82.064010}.
Individually, we choose one parameter from $\psi_0$, $\psi_2$ and $\psi_{6l}$ and let it freely vary together with BBH parameters. Based on the parameterized IMRPhenomXAS waveform model we can get the posterior distribution of the chosen $\psi_{j(l)}$ parameter and BBH parameters. For each value located in the posterior distribution region of the chosen $\psi_{j(l)}$, one curve is determined in the mass--mass diagram according to the functional form of this $\psi_{j(l)}$ in GR theory. Running over the whole posterior distribution region of the chosen $\psi_{j(l)}$, a corresponding region in the mass--mass diagram is determined. For reference convenience, we call this region the GR possible region. On the other hand, the posterior distribution of $m_1$-$m_2$ also gives a region. Then the aforementioned GR possible region should intersect with the posterior distribution of $m_1$-$m_2$ according to GR theory. If they do not intersect, GR theory is ruled out.

The associated credible level of the above-mentioned posterior distribution region corresponds to the level at which the data show a tension with GR (ruling GR out) or are consistent with GR (ruling GR in).
We note that this critical probability is a Bayesian credible-level construct obtained from posterior samples; its interpretation is conditional on the adopted priors, waveform model, and the definition of the accumulated probability regions.

In our implementation, the inverse mapping is performed numerically in \textsc{Mathematica}; the posterior intervals for the PN coefficient are taken as equal-tail intervals, motivated by the approximate symmetry of the $\delta_{j(l),\text{NS}}$ posteriors in the LVK results; and the probability densities are obtained via kernel density estimation from the posterior samples.

\section{Null test of GR based on O1 and O2 BBH events}\label{sec3}
We use the \texttt{pycbc\_inference} executable \cite{Biwer:2018osg} based on \texttt{PyCBC} \cite{alex_nitz_2024_10473621} to obtain the posteriors of parameter estimation. We modify IMRPhenomXAS based on the purely Python library \texttt{WF4Py} \cite{Iacovelli_WF4Py_2022,Iacovelli:2022bbs,Iacovelli:2022mbg} to implement our parameterized IMRPhenomXAS waveform model.

The settings of \texttt{pycbc\_inference} \texttt{***.ini} files for the ten BBH events of O1 and O2 (for which we perform parameter estimation as a validation of our parameterized IMRPhenomXAS waveform) are based on the configuration files proposed by 4-OGC \cite{2023ApJ...946...59N} including the priors of every parameter, model settings and sampler settings. Specifically we set the mass ratio $1<q<10$ and the chirp mass $10M_\odot<\mathcal{M}<100M_\odot$.

The boundaries of uniform priors of all PN coefficients are set to be $10$ times greater or less than the maximum or minimum of GR PN coefficient values over the GR parameters range $m_1, m_2 \in [1,100]$ and $\chi_{1L}, \chi_{2L} \in [-0.99,0.99]$. Specifically we have $-15000<\psi_0<15000$, $-10000<\psi_2<10000$, and $-20<\psi_{6l}<20$. All PyCBC PE configuration files including the modifications mentioned above are available on GitHub\footnote{https://github.com/chendawe/parameterized-test-local.git}. These analyses largely follow the workflow of the LVK collaboration's existing study \cite{PhysRevD.100.104036}. We find our results are consistent with those presented by the LVK collaboration.
Meanwhile, the LVK collaboration's published parameterized test results for spin-independent PN coefficients are available for five of these ten events (GW150914, GW151226, GW170104, GW170608, and GW170814); our null-test analysis in the remainder of this paper is based on these five events.
\begin{figure*}
\begin{tabular}{cc}
\includegraphics[width=0.45\textwidth]{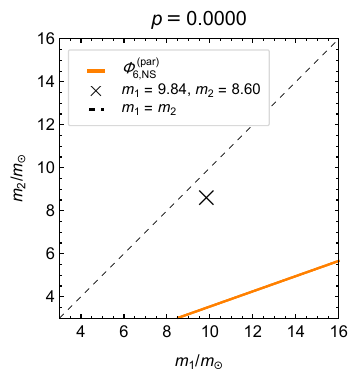} &
\includegraphics[width=0.45\textwidth]{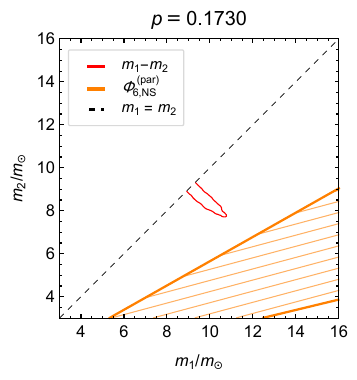} \\
\makebox[0.45\textwidth][c]{(a)} &
\makebox[0.45\textwidth][c]{(b)} \\[6pt]
\includegraphics[width=0.45\textwidth]{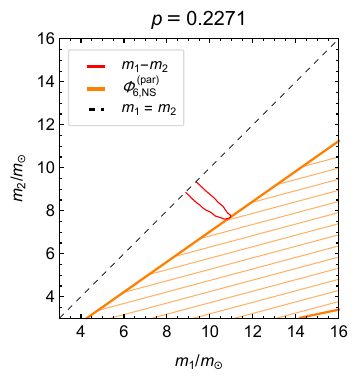} &
\includegraphics[width=0.45\textwidth]{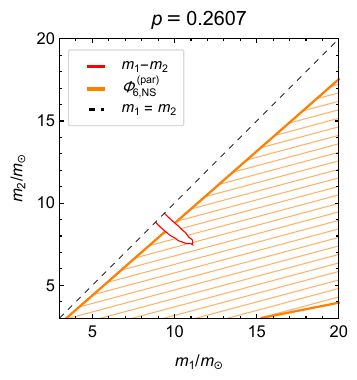} \\
\makebox[0.45\textwidth][c]{(c)} &
\makebox[0.45\textwidth][c]{(d)}
\end{tabular}
\caption{Null test scheme of GR with a single coefficient based on LVK analysis of GW170608 for $\phi_{6,\text{NS}}$. Starting from the most probable point and line, the accumulated probability increases from top to bottom and from left to right. The left-bottom subplot corresponds to the kissing situation and we call it critical case.
At the critical probability, the GR-consistent regions first intersect, corresponding to a tension with GR at credible level $22.71$\% (equivalently, GR consistency at credible level $77.29$\%).
}\label{fig1}
\end{figure*}

Before proceeding, we clarify the notation conventions used in the remainder of this paper. In Sec.~\ref{sec2}, the full PN coefficient $\psi_{j(l)}$ includes the common mass-dependent prefactor $\text{Coeff}(M,\eta)=\frac{3}{128}\frac{1}{\eta}(\pi M)^{(j-5)/3}$, i.e., $\psi_{j(l)}=\text{Coeff}(M,\eta)\cdot\alpha_{j(l)}$, where $\alpha_{j(l)}$ is the dimensionless part. In contrast, the LVK collaboration's parameterized IMRPhenomPv2 analysis reports the fractional deviation $\delta_{j(l),\text{NS}}$ defined with respect to the dimensionless coefficient $\phi_{j(l),\text{NS}}$. We transform the LVK $\delta$ posteriors into dimensionless PN-coefficient posteriors via $\phi_{j(l),\text{NS}}=\phi^{\rm GR}_{j(l),\text{NS}}(1+\delta_{j(l),\text{NS}})$. Our null-test framework is independent of which convention is adopted, because the inverse mapping from a measured coefficient to a mass--mass relation via its GR functional form is universal. We note that the two conventions also differ in the treatment of the $\ln(\pi M f)$ term: in the $\psi$-convention it is split into $\ln(\pi M)+\ln f$ (absorbed into $\psi_j$ and $\psi_{jl}$ respectively), while in the dimensionless $\phi$-convention the entire factor is assigned to the log-order coefficient. Since the LVK data use the dimensionless convention, we adopt $\phi_{j(l),\text{NS}}$ throughout Sec.~\ref{sec3}.

In order to avoid redundancy and confusion, in the remainder of current paper we proceed based on the LVK collaboration's results of $\delta_{j(l),\text{NS}}$ \cite{PhysRevD.100.104036}. More specifically the LVK collaboration used the parameterized IMRPhenomPv2 waveform model to analyze the BBH events included in GWTC-1. The involved spin-independent PN coefficients include
\begin{align}
    \phi_{2,\text{NS}} &= \frac{3715}{756} + \frac{55}{9}\eta,\\
    \phi_{4,\text{NS}} &= \frac{15293365}{508032} + \frac{27145}{504}\eta + \frac{3085}{72}\eta^2,\\
    \phi_{5l,\text{NS}} &= \frac{5\pi}{3}\left(\frac{7729}{84} - 13\eta\right),\\
    \phi_{6,\text{NS}} &= \frac{11583231236531}{4694215680} - \frac{640}{3}\pi^{2} - \frac{6848}{21}\gamma_\text{E} \notag\\
    &\quad + \eta\left(-\frac{15737765635}{3048192} + \frac{2255}{12}\pi^{2}\right) + \frac{76055}{1728}\eta^{2} \notag\\
    &\quad - \frac{127825}{1296}\eta^{3} - \frac{6848}{21}\ln 4,\\
    \phi_{7,\text{NS}} &= \pi\left(\frac{77096675}{254016} + \frac{378515}{1512}\eta - \frac{74045}{756}\eta^{2}\right)
\end{align}
where $\gamma_\text{E}$ is the Euler constant. By performing the similar transformation $\phi_{j(l),\text{NS}} = \phi^\text{GR}_{j(l),\text{NS}} \cdot (1 + \delta_{j(l),\text{NS}})$ as described in the last section for $\psi_{j(l)}$, we use the final results of $\phi_{j(l),\text{NS}}$ to do the null test of GR based on O1 and O2 BBH events.

Furthermore, we evaluate the critical probabilities of multi-parameter cases, which are the credible levels at which the GR mass region and the GR possible regions of multiple parameters first share a common intersection.
This is a direct generalization of the definition of single-parameter critical probability. We can therefore consider the overall consistencies of GR with multiple parameters simultaneously.

\begin{table}
  \centering
  \caption{The critical probability at which the GR-consistent regions first intersect, based on LVK data analysis results for O1/O2 BBH events. Together with these critical probabilities, we also list the network SNR corresponding to each event.}  \label{tab1}
  \begin{tabular}
    {|p{1.7cm}<\centering |p{0.8cm}<\centering |p{1cm}<\centering |p{1cm}<\centering |p{1cm}<\centering|p{1cm}<\centering | p{1cm}<\centering|}
    \hline
    BBH ID & SNR & $\phi_{2,\text{NS}}$ & $\phi_{4,\text{NS}}$ & $\phi_{5l,\text{NS}}$ & $\phi_{6,\text{NS}}$ & $\phi_{7,\text{NS}}$ \\ \hline
    GW150914 & $26.0$ & $0.9225$ & $0.9309$ & $0.9559$ & $0.9332$ & $0.9569$ \\
    GW151226 & $13.1$ & $0.0974$ & $0.0334$ & $0.0670$ & $0.1138$ & $0.1589$ \\
    GW170104 & $13.8$ & $0.9243$ & $0.9124$ & $0.9316$ & $0.9017$ & $0.9060$ \\
    GW170608 & $15.4$ & $0.4238$ & $0.3454$ & $0.2874$ & $0.2271$ & $0.1464$ \\
    GW170814 & $17.7$ & $0.5834$ & $0.3411$ & $0.1923$ & $0.1194$ & $0.0568$ \\
    \hline
  \end{tabular}
\end{table}

\subsection{Null test of GR with a single coefficient}
We use $\phi_{6,\text{NS}}$ of GW170608 as an example to illustrate the test scheme we proposed in Sec.~\ref{secIIC}. In Fig.~\ref{fig1}a, we plot the most probable point based on GR waveform analysis and the most probable line based on the extended waveform of $\phi_{6,\text{NS}}$ analysis in $m_1$-$m_2$ figure. The point does not lie on the line, which is different from the prediction of GR. In this sense the data show a tension with GR, and formally the accumulated probability is zero. Then we increase the accumulated probability, the point and the line become two regions as illustrated in Fig.~\ref{fig1}b which corresponds to the probability $p=0.1730$. Since the two regions do not touch, the data show a tension with GR at a credible level above 0.173. We increase the accumulated probability more to $p=0.2271$ to get two kissing regions as plotted in Fig.~\ref{fig1}c.
We call this probability the critical probability: at $p=0.2271$, the data show a tension with GR at credible level $22.71$\%, or equivalently, GR is consistent with the data at credible level 77.29\%.

LVK has analyzed GW150914, GW151226, GW170104, GW170608 and GW170814 for $\phi_{2,\text{NS}}$, $\phi_{4,\text{NS}}$, $\phi_{5l,\text{NS}}$, $\phi_{6,\text{NS}}$ and $\phi_{7,\text{NS}}$ in \cite{PhysRevD.100.104036}. We list the corresponding critical probability in Tab.~\ref{tab1}.
We find that GW150914 and GW170104 exhibit a high credible level (more than about 90\%) for tension with GR, while the other three events admit much smaller credible levels (less than about 60\%).
At the same time, we can note that all $\phi_{j(l),\text{NS}}$ admit consistent critical probabilities.

Notably, we find no monotonic correlation between the network SNR and the critical probability for tension with GR: GW150914 (SNR$=26.0$) and GW170104 (SNR$=13.8$) both show high tension credible levels ($\gtrsim 90\%$), while GW170814 (SNR$=17.7$) shows a much lower level. This suggests that the observed tensions are not simply governed by signal strength and may originate from waveform systematics or event-specific physical properties.

\begin{figure*}
\begin{tabular}{cc}
\includegraphics[width=0.45\textwidth]{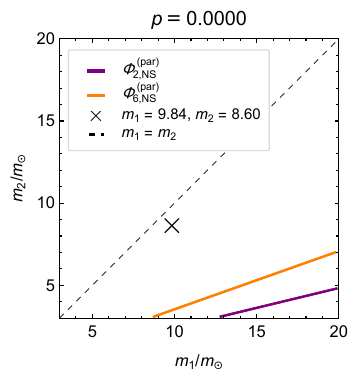} &
\includegraphics[width=0.45\textwidth]{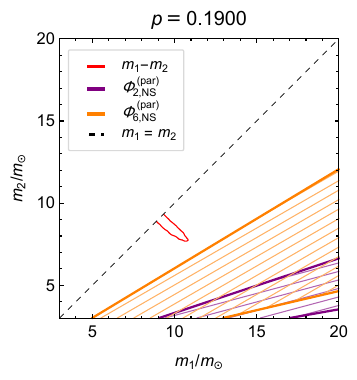} \\
\makebox[0.45\textwidth][c]{(a)} &
\makebox[0.45\textwidth][c]{(b)} \\[6pt]
\includegraphics[width=0.45\textwidth]{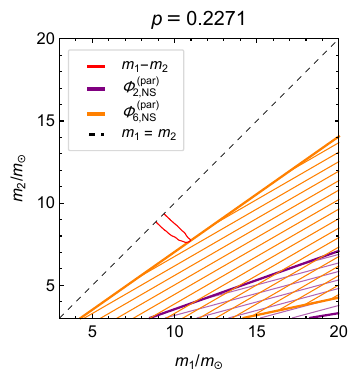} &
\includegraphics[width=0.45\textwidth]{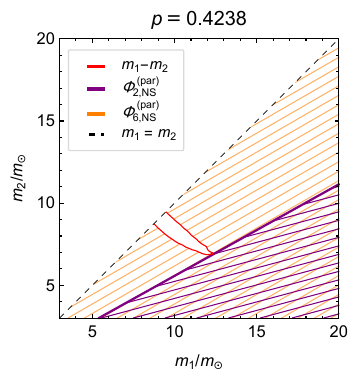} \\
\makebox[0.45\textwidth][c]{(c)} &
\makebox[0.45\textwidth][c]{(d)}
\end{tabular}
\caption{Null test scheme of GR with two coefficients based on LVK analysis of GW170608 for $\phi_{2,\text{NS}}$ and $\phi_{6,\text{NS}}$. Starting from the most probable point and lines, the accumulated probability increases from top to bottom and from left to right. The right-bottom subplot corresponds to the common intersection of the three regions and we call it critical case.
At the critical probability, the GR-consistent regions first intersect, corresponding to a tension with GR at credible level $42.38$\% (equivalently, GR consistency at credible level $57.62$\%).
}\label{fig2}
\end{figure*}

\begin{figure*}
\begin{tabular}{cc}
\includegraphics[width=0.45\textwidth]{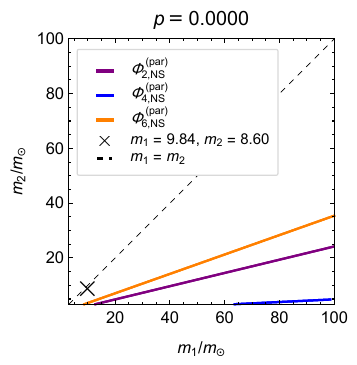} &
\includegraphics[width=0.45\textwidth]{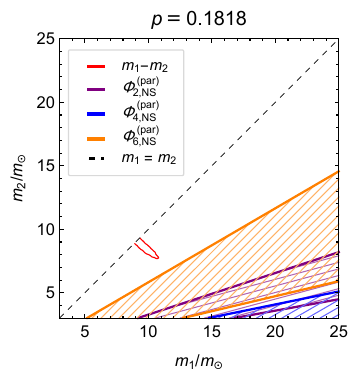} \\
\makebox[0.45\textwidth][c]{(a)} &
\makebox[0.45\textwidth][c]{(b)} \\[6pt]
\includegraphics[width=0.45\textwidth]{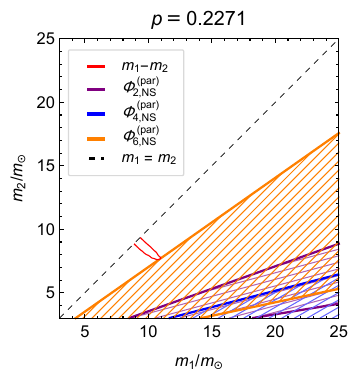} &
\includegraphics[width=0.45\textwidth]{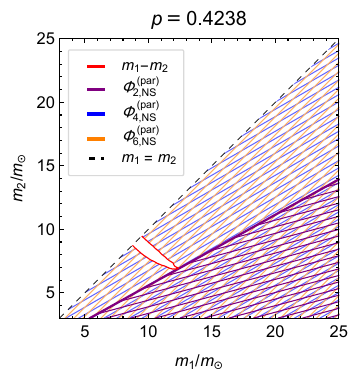} \\
\makebox[0.45\textwidth][c]{(c)} &
\makebox[0.45\textwidth][c]{(d)}
\end{tabular}
\caption{Null test scheme of GR with three coefficients based on LVK analysis of GW170608 for $\phi_{2,\text{NS}}$, $\phi_{4,\text{NS}}$ and $\phi_{6,\text{NS}}$. Starting from the most probable point and lines, the accumulated probability increases from top to bottom and from left to right. The right-bottom subplot corresponds to the common intersection of the four regions and we call it critical case.
At the critical probability, the GR-consistent regions first intersect, corresponding to a tension with GR at credible level $42.38$\% (equivalently, GR consistency at credible level $57.62$\%).
}\label{fig3}
\end{figure*}

\begin{figure*}
\begin{tabular}{cc}
\includegraphics[width=0.45\textwidth]{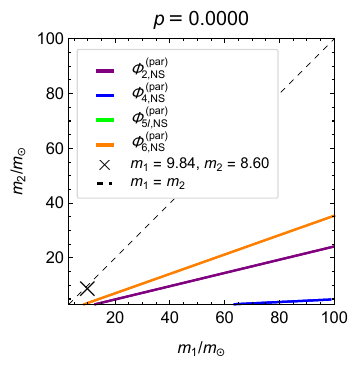} &
\includegraphics[width=0.45\textwidth]{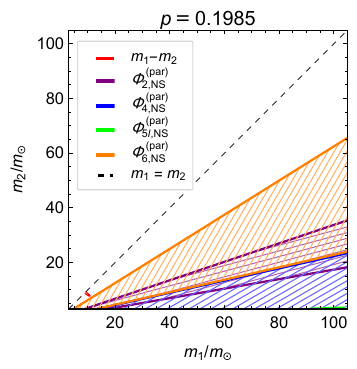} \\
\makebox[0.45\textwidth][c]{(a)} &
\makebox[0.45\textwidth][c]{(b)} \\[6pt]
\includegraphics[width=0.45\textwidth]{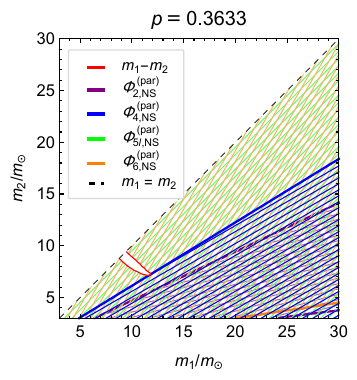} &
\includegraphics[width=0.45\textwidth]{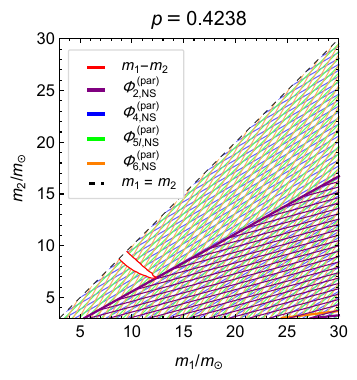} \\
\makebox[0.45\textwidth][c]{(c)} &
\makebox[0.45\textwidth][c]{(d)}
\end{tabular}
\caption{Null test scheme of GR with four coefficients based on LVK analysis of GW170608 for $\phi_{2,\text{NS}}$, $\phi_{4,\text{NS}}$, $\phi_{5l,\text{NS}}$ and $\phi_{6,\text{NS}}$. Starting from the most probable point and lines, the accumulated probability increases from top to bottom and from left to right. The right-bottom subplot corresponds to the common intersection of the five regions and we call it critical case.
At the critical probability, the GR-consistent regions first intersect, corresponding to a tension with GR at credible level $42.38$\% (equivalently, GR consistency at credible level $57.62$\%).
}\label{fig4}
\end{figure*}

\subsection{Null test of GR with two coefficients}
To demonstrate the two-coefficient test, we take $\phi_{2,\text{NS}}$ and $\phi_{6,\text{NS}}$ of GW170608 and follow the same accumulated-probability procedure described in Sec.~\ref{secIIC}.
Figure \ref{fig2} displays the results: the GR point and the two most probable lines from the extended waveforms initially show no common intersection. As the probability increases, the point and lines expand into regions (Figs.~\ref{fig2}b and \ref{fig2}c). A single common intersection shared by all three regions first appears at an accumulated probability $p = 0.4238$ (Fig.~\ref{fig2}d).
We take this value as the critical probability for the two-coefficient test, at which the data show a tension with GR at credible level 42.38\%.

Similar to Tab.~\ref{tab1}, we list the corresponding critical probability for two coefficients test in Tab.~\ref{tab2} for GW150914, GW151226, GW170104, GW170608 and GW170814.
\begin{table*}[htbp]
  \centering
  \caption{The critical probability at which the GR-consistent regions first intersect, based on LVK data analysis results for O1/O2 BBH events, for combinations of two parameters.
  The probability for each combination is the critical credible level at which the regions of the two coefficients and the mass region first intersect.
  The network SNR for each event is also listed.}
  \label{tab2}
  \begin{tabular}
    {|p{1.7cm}<\centering |p{0.8cm}<\centering |p{1.3cm}<\centering |p{1.3cm}<\centering |p{1.3cm}<\centering|p{1.3cm}<\centering | p{1.3cm}<\centering|p{1.3cm}<\centering |p{1.3cm}<\centering |p{1.3cm}<\centering|p{1.3cm}<\centering | p{1.3cm}<\centering|}
    \hline
    BBH ID & SNR
    & $\phi_{2,4,\text{NS}}$ & $\phi_{2,5l,\text{NS}}$ & $\phi_{2,6,\text{NS}}$ & $\phi_{2,7,\text{NS}}$ & $\phi_{4,5l,\text{NS}}$ & $\phi_{4,6,\text{NS}}$& $\phi_{4,7,\text{NS}}$& $\phi_{5l,6,\text{NS}}$
    & $\phi_{5l,7,\text{NS}}$& $\phi_{6,7,\text{NS}}$\\
    \hline
    GW150914 & $26.0$ & $0.9309$ & $0.9559$ & $0.9332$ & $0.9569$ & $0.9559$ & $0.9332$ & $0.9569$ & $0.9559$ & $0.9569$ & $0.9569$ \\
    GW151226 & $13.1$ & $0.0974$ & $0.0974$ & $0.1138$ & $0.1589$ & $0.0670$ & $0.1138$ & $0.1589$ & $0.1138$ & $0.1589$ & $0.1589$ \\
    GW170104 & $13.8$ & $0.9243$ & $0.9316$ & $0.9243$ & $0.9243$ & $0.9316$ & $0.9124$ & $0.9124$ & $0.9316$ & $0.9316$ & $0.9060$ \\
    GW170608 & $15.4$ & $0.4238$ & $0.4238$ & $0.4238$ & $0.4238$ & $0.3454$ & $0.3454$ & $0.3454$ & $0.2874$ & $0.2874$ & $0.2271$ \\
    GW170814 & $17.7$ & $0.5834$ & $0.5834$ & $0.5834$ & $0.5834$ & $0.3411$ & $0.3411$ & $0.3411$ & $0.1923$ & $0.1923$ & $0.1194$ \\
    \hline
  \end{tabular}
\end{table*}

Before proceeding to higher-order combinations, we comment on a notable feature visible in Tab.~\ref{tab2}: several multi-parameter critical credible probabilities are identical to one of the corresponding single-parameter values. This is a mathematical consequence of the construction, not a numerical artifact. The multi-parameter critical probability is defined as the smallest accumulated probability $p$ at which the mass posterior region and all the coefficient-specific GR-possible regions share a common intersection. If one coefficient's GR-possible region is entirely contained within (or exactly tangent to) the other regions at the critical point, the common intersection is determined solely by the most restrictive region---the one requiring the largest accumulated probability to intersect the mass posterior. Adding more coefficients then does not change the critical probability, because the additional regions do not further constrain the common intersection. For example, in GW170608 the $\phi_{2,\text{NS}}$ region is the most restrictive (single-parameter critical probability $0.4238$), so any multi-parameter combination including $\phi_{2,\text{NS}}$ yields $0.4238$; similarly, for GW150914 the $\phi_{7,\text{NS}}$ coefficient ($0.9569$) is dominant. The same principle applies to the three-, four-, and five-coefficient tests below.

\subsection{Null test of GR with three coefficients}
We apply the three coefficient test to GW170608 using $\phi_{2,\text{NS}}$, $\phi_{4,\text{NS}}$, and $\phi_{6,\text{NS}}$, following the same accumulated probability procedure described in Sec.~\ref{secIIC}.
Figure \ref{fig3} displays the results: the GR point and the three most probable lines expand into four regions as the probability increases (Figs.~\ref{fig3}b and \ref{fig3}c). A common intersection shared by all four regions first appears at $p = 0.4238$ (Fig.~\ref{fig3}d).
We take this value as the critical probability for the three-coefficient test, at which the data show a tension with GR at credible level 42.38\%.

The critical probabilities obtained from all three-coefficient combinations for the five O1/O2 BBH events are listed in Table \ref{tab3}.
\begin{table*}[htbp]
  \centering
  \caption{The critical probability at which the GR-consistent regions first intersect, based on LVK data analysis results for O1/O2 BBH events, for combinations of three parameters.
  The probability for each combination is the critical credible level at which the regions of the three coefficients and the mass region first intersect.
  The network SNR for each event is also listed.}
  \label{tab3}
  \begin{tabular}
    {|p{1.7cm}<\centering |p{0.8cm}<\centering |p{1.3cm}<\centering |p{1.3cm}<\centering |p{1.3cm}<\centering|p{1.3cm}<\centering | p{1.3cm}<\centering|p{1.3cm}<\centering |p{1.3cm}<\centering |p{1.3cm}<\centering|p{1.3cm}<\centering | p{1.3cm}<\centering|}
    \hline
    BBH ID & SNR
    & $\phi_{2,4,5l,\text{NS}}$ & $\phi_{2,4,6,\text{NS}}$ & $\phi_{2,4,7,\text{NS}}$ & $\phi_{2,5l,6,\text{NS}}$ & $\phi_{2,5l,7,\text{NS}}$ & $\phi_{2,6,7,\text{NS}}$ & $\phi_{4,5l,6,\text{NS}}$ & $\phi_{4,5l,7,\text{NS}}$
    & $\phi_{4,6,7,\text{NS}}$ & $\phi_{5l,6,7,\text{NS}}$\\
    \hline
    GW150914 & $26.0$ & $0.9559$ & $0.9332$ & $0.9569$ & $0.9559$ & $0.9569$ & $0.9569$ & $0.9559$ & $0.9569$ & $0.9569$ & $0.9569$ \\
    GW151226 & $13.1$ & $0.0974$ & $0.1138$ & $0.1589$ & $0.1138$ & $0.1589$ & $0.1589$ & $0.1138$ & $0.1589$ & $0.1589$ & $0.1589$ \\
    GW170104 & $13.8$ & $0.9316$ & $0.9243$ & $0.9243$ & $0.9316$ & $0.9316$ & $0.9243$ & $0.9316$ & $0.9316$ & $0.9124$ & $0.9316$ \\
    GW170608 & $15.4$ & $0.4238$ & $0.4238$ & $0.4238$ & $0.4238$ & $0.4238$ & $0.4238$ & $0.3454$ & $0.3454$ & $0.3454$ & $0.2874$ \\
    GW170814 & $17.7$ & $0.5834$ & $0.5834$ & $0.5834$ & $0.5834$ & $0.5834$ & $0.5834$ & $0.3411$ & $0.3411$ & $0.3411$ & $0.1923$ \\
    \hline
  \end{tabular}
\end{table*}
\subsection{Null test of GR with four coefficients}
We now combine the deviations of four PN coefficients together and use $\phi_{2,\text{NS}}$, $\phi_{4,\text{NS}}$, $\phi_{5l,\text{NS}}$, and $\phi_{6,\text{NS}}$ of GW170608 as an illustration.
As shown in Fig.~\ref{fig4}, the line corresponding to $\phi_{5l,\text{NS}}$ is initially absent at the most probable value, because no physical ($m_1,m_2$) pair can produce that deviation within the mass range $[10,100]\,M_\odot$. The line appears only when the accumulated probability reaches $p=0.1985$ (Fig.~\ref{fig4}b). The GR point and the four lines eventually form five regions, which first share a common intersection at $p = 0.4238$ (Fig.~\ref{fig4}d).
We take this value as the critical probability for the four-coefficient test, at which the data show a tension with GR at credible level 42.38\%.

The results for all four parameter configurations are collected in Table \ref{tab4}.
\begin{table*}[htbp]
  \centering
  \caption{The critical probability at which the GR-consistent regions first intersect, based on LVK data analysis results for O1/O2 BBH events, for combinations of four parameters.
  The probability for each combination is the critical credible level at which the regions of the four coefficients and the mass region first intersect.
  The network SNR for each event is also listed.}
  \label{tab4}
  \begin{tabular}
    {|p{1.9cm}<\centering |p{1cm}<\centering |p{1.6cm}<\centering |p{1.6cm}<\centering |p{1.6cm}<\centering|p{1.6cm}<\centering | p{1.6cm}<\centering|}
    \hline
    BBH ID & SNR
    & $\phi_{2,4,5l,6,\text{NS}}$ & $\phi_{2,4,5l,7,\text{NS}}$ & $\phi_{2,4,6,7,\text{NS}}$ & $\phi_{2,5l,6,7,\text{NS}}$ & $\phi_{4,5l,6,7,\text{NS}}$ \\
    \hline
    GW150914 & $26.0$ & $0.9559$ & $0.9569$ & $0.9569$ & $0.9569$ & $0.9569$ \\
    GW151226 & $13.1$ & $0.1138$ & $0.1589$ & $0.1589$ & $0.1589$ & $0.1589$ \\
    GW170104 & $13.8$ & $0.9316$ & $0.9316$ & $0.9243$ & $0.9316$ & $0.9316$ \\
    GW170608 & $15.4$ & $0.4238$ & $0.4238$ & $0.4238$ & $0.4238$ & $0.3454$ \\
    GW170814 & $17.7$ & $0.5834$ & $0.5834$ & $0.5834$ & $0.5834$ & $0.3411$ \\
    \hline
  \end{tabular}
\end{table*}

\begin{figure*}
\begin{tabular}{cc}
\includegraphics[width=0.45\textwidth]{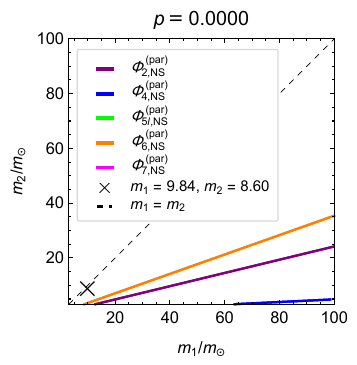} &
\includegraphics[width=0.45\textwidth]{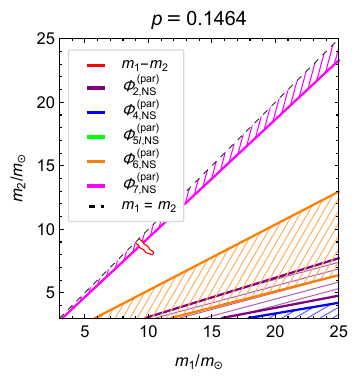} \\
\makebox[0.45\textwidth][c]{(a)} &
\makebox[0.45\textwidth][c]{(b)} \\[6pt]
\includegraphics[width=0.45\textwidth]{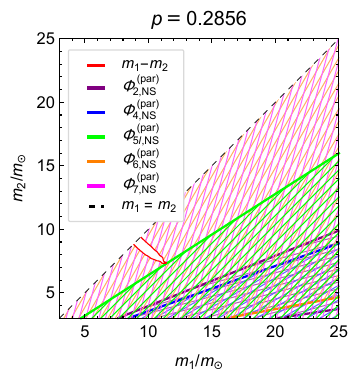} &
\includegraphics[width=0.45\textwidth]{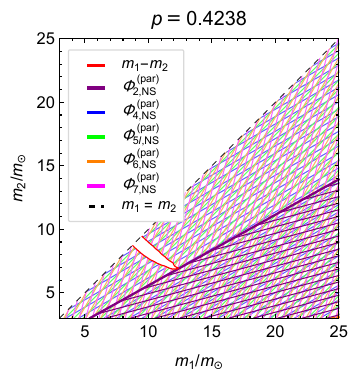} \\
\makebox[0.45\textwidth][c]{(c)} &
\makebox[0.45\textwidth][c]{(d)}
\end{tabular}
\caption{Null test scheme of GR with five coefficients based on LVK analysis of GW170608 for $\phi_{2,\text{NS}}$, $\phi_{4,\text{NS}}$, $\phi_{5l,\text{NS}}$, $\phi_{6,\text{NS}}$ and $\phi_{7,\text{NS}}$. Starting from the most probable point and lines, the accumulated probability increases from top to bottom and from left to right. The right-bottom subplot corresponds to the common intersection of the six regions and we call it critical case.
At the critical probability, the GR-consistent regions first intersect, corresponding to a tension with GR at credible level $42.38$\% (equivalently, GR consistency at credible level $57.62$\%).
}\label{fig5}
\end{figure*}
\subsection{Null test of GR with five coefficients}
Finally, we test the full set of five spin-independent PN coefficients: $\phi_{2,\text{NS}}$, $\phi_{4,\text{NS}}$, $\phi_{5l,\text{NS}}$, $\phi_{6,\text{NS}}$, and $\phi_{7,\text{NS}}$. Figure \ref{fig5} shows the results for GW170608. Similar to the four coefficient case, the lines for $\phi_{5l,\text{NS}}$ and $\phi_{7,\text{NS}}$ are missing at the most probable value and become visible only at higher probabilities ($p=0.1985$ and $p=0.1464$, respectively). The six regions first intersect at $p = 0.4238$ (Fig.~\ref{fig5}d),
giving a critical probability of 42.38\% at which the data show a tension with GR.

Table \ref{tab5} lists the critical probabilities for all five events.
\begin{table}[htbp]
  \centering
  \caption{The critical probability at which the GR-consistent regions first intersect, based on LVK data analysis results for O1/O2 BBH events, for the combination of all five parameters. The probability is the critical credible level at which the regions of the five coefficients and the mass region first intersect. The network SNR for each event is also listed.}
  \label{tab5}
    \begin{tabular}
    {|p{1.9cm}<\centering |p{1cm}<\centering |p{3cm}<\centering |}
    \hline
    BBH ID & SNR
    & $\phi_{2,4,5l,6,7,\text{NS}}$ \\
    \hline
    GW150914 & $26.0$ & $0.9569$ \\
    GW151226 & $13.1$ & $0.1589$ \\
    GW170104 & $13.8$ & $0.9316$ \\
    GW170608 & $15.4$ & $0.4238$ \\
    GW170814 & $17.7$ & $0.5834$ \\
    \hline
  \end{tabular}
\end{table}

\section{Conclusion and discussion}
Gravitational waves are an ideal tool to test GR. Based on gravitational wave observation data, many test methods have been proposed. Among these methods, the parameterized scheme is one of the most extensively studied. Several LVK reports on GR test are based on such scheme.

Although the parameterized scheme is very powerful for testing GR, there is a critical issue involved. Based on the test results, one can claim that GR is preferred by the data if we find the deviation parameter is small. If the deviation parameters are closer to zero, one can say that the data are more consistent with GR. But the problem is how small is small enough?

In the current paper, we propose a null test scheme that extends the parameterized scheme by one additional step. In fact our idea is not completely new; it is an extension of existing works \cite{PhysRevD.74.024006,2006CQGra..23L..37A,PhysRevD.82.064010}. The existing works only consider the inspiral stage of CBC waves. Here we extend to full inspiral-merger-ringdown waves which can use the whole detection data by LVK to avoid the uncertainty of inspiral stage cutting.

Borrowing the idea of mass--mass relations used in the pulsar binary studies, we transform the analysis results of parameterized test scheme into mass--mass relations. Through comparing the resulting mass--mass relations coming from different parameters, we can estimate the probabilities that the data prefer the GR theory. In this way we answered the question how small a deviation parameter's value is small enough.

In the current paper we have analyzed five O1/O2 LVK BBH events. Among them, GW150914 and GW170104 show a tension with GR at a credible level above 90\%. In contrast, GW151226, GW170608 and GW170814 are more consistent with GR. Moreover, we find no monotonic correlation between the network SNR and the credible level for tension with GR, suggesting that the observed tensions are not simply driven by signal strength but may reflect waveform systematics or event-specific properties. Based on these facts, we conclude that current gravitational wave observation data may include some information deviating GR.
We caution that the observed deviations should not be interpreted as evidence for a breakdown of GR. They may arise from several systematic sources: (i) waveform model systematics, including the known limitations of IMRPhenomPv2 in the merger--ringdown regime and differences between waveform families; (ii) detector noise artifacts and calibration uncertainties, which can bias parameter estimation especially for lower-SNR events; (iii) prior choices, particularly the wide uniform priors on the PN deviation parameters, which can produce broad posteriors that overlap zero only at large credible levels; and (iv) parameter correlations, since the PN coefficients are degenerate with source parameters such as the spins and mass ratio. Our null test should therefore be viewed as a complementary diagnostic that motivates further investigation with more accurate waveform models, independent analysis pipelines, and a larger catalog of events, rather than as conclusive evidence against GR.

Our null test scheme can be generalized to other waveform models straightforwardly. And our test scheme can be easily applied to all of the GWTC-1 to -5 data sets \cite{LIGOScientific:2026sit}. Our null test scheme provides a good complement to the existing GR test schemes based on gravitational waves.

\section*{Acknowledgments}
This work was supported in part by the National Key Research and Development Program of China Grant No. 2021YFC2203001 and in part by the NSFC (No.~12475046).
\begin{appendix}

\section{Explicit Expressions of Post-Newtonian Coefficients}
\label{appendix:pn_coefficient_analytic_expressions}
In this appendix section we list the PN coefficients appearing in (\ref{eq3}) and (\ref{eq4}) for GR theory. The non-spin parts of integer-order-type PN coefficients
\begin{align}
&   \alpha_{0,\text{NS}} = 1.0, \\
&   \alpha_{1,\text{NS}} = 0.0, \\
&   \alpha_{2,\text{NS}} = \frac{3715}{756} + \frac{55}{9}\eta, \\
&   \alpha_{3,\text{NS}} = -16 \pi, \\
&   \alpha_{4,\text{NS}} = \frac{15293365}{508032} + \frac{27145}{504}\eta + \frac{3085}{72}\eta^2, \\
&   \alpha_{5,\text{NS}} = 0.0, \\
&   \alpha_{6,\text{NS}} = \frac{11583231236531}{4694215680} - \frac{640\pi^2}{3} + \frac{2255\eta\pi^2}{12}\nonumber \\
                        &- \frac{5}{3048192}\left[3147553127 + 588\eta \left(-45633 + 102260\eta\right)\right]\eta\nonumber\\
                        &- \frac{6848\gamma_\text{E}}{21} - \frac{13696\ln 2}{21} - \frac{6848\ln \pi}{63}, \\
&   \alpha_{7,\text{NS}} = \frac{5\pi}{254016} \Bigl[ 15419335 + 168\,(75703 - 29618\,\eta)\,\eta \Bigr],
\end{align}
where $\gamma_\text{E}$ is the Euler constant.

The spin parts of integer-order-type PN coefficients
\begin{align}
&\alpha_{0,\text{S}} = 0.0, \\
&\alpha_{1,\text{S}} = 0.0, \\
&\alpha_{2,\text{S}} = 0.0, \\
&\alpha_{3,\text{S}} = \frac{113}{6} \left[\left( \chi_{1L} + \chi_{2L} + \chi_{1L}\delta - \chi_{2L}\delta \right)\right.\nonumber\\
&\left.- 76 \left( \chi_{1L} + \chi_{2L} \right)\eta\right], \\
&\alpha_{4,\text{S}} = \frac{-5}{16} \left[ 81 \chi_{1L}^2 (1 + \delta - 2\eta) + 316 \chi_{1L}\chi_{2L}\eta \right.\nonumber\\
&\left.- 81 \chi_{2L}^2 (-1 + \delta + 2\eta) \right], \\
&\alpha_{5,\text{S}} = 0.0, \\
&\alpha_{6,\text{S}} = \frac{5}{3}\pi \left[ 227 \left( \chi_{1L} + \chi_{2L} + \chi_{1L}\delta - \chi_{2L}\delta \right)\right.\nonumber\\
&\left.- 156 \left( \chi_{1L} + \chi_{2L} \right)\eta \right], \\
&\alpha_{7,\text{S}} = \frac{5}{6096384} \Big[ -5030016755 \left( \chi_{1L} + \chi_{2L} + \chi_{1L}\delta - \chi_{2L}\delta \right) \nonumber\\
&+ 4 \left( 2113331119(\chi_{1L} + \chi_{2L}) + 675484362(\chi_{1L} - \chi_{2L})\delta \right)\eta \nonumber\\
&- 1008 \left( 208433(\chi_{1L} + \chi_{2L}) + 25011(\chi_{1L} - \chi_{2L})\delta \right)\eta^2 \nonumber\\
&+ 90514368(\chi_{1L} + \chi_{2L})\eta^3 \Big] - 5 \left[ 57 \chi_{1L}^2 (1 + \delta - 2\eta) \right.\nonumber\\
&\left.+ 220 \chi_{1L}\chi_{2L}\eta - 57 \chi_{2L}^2 (-1 + \delta + 2\eta) \right] \nonumber\\
&+ \frac{14585}{48} \Big[\left( \chi_{1L}^3(1 + \delta) - \chi_{2L}^3(-1 + \delta) \right) \nonumber\\
&- 5 \left[ 8819\chi_{2L}^3 + 8439\chi_{1L}\chi_{2L}^2(-1 + \delta)\right.\nonumber\\
&\left. - 8439\chi_{1L}^2\chi_{2L}(1 + \delta) + \chi_{1L}^3(8819 + 2985\delta) \right]\eta\nonumber \\
&+ 40(\chi_{1L} + \chi_{2L})(17\chi_{1L}^2 - 14\chi_{1L}\chi_{2L} + 17\chi_{2L}^2)\eta^2\Big].
\end{align}
Here $\delta\equiv|m_1 - m_2| / (m_1 + m_2)$ is the relative mass difference between the two black holes.

The non-spin parts of log-order-type PN coefficients
\begin{flalign}
    \alpha_{0l,\text{NS}} &= 0.0, \\
    \alpha_{1l,\text{NS}} &= 0.0, \\
    \alpha_{2l,\text{NS}} &= 0.0, \\
    \alpha_{3l,\text{NS}} &= 0.0, \\
    \alpha_{4l,\text{NS}} &= 0.0, \\
    \alpha_{5l,\text{NS}} &= \frac{5}{4536}\pi(46374 - 6552 \eta), \\
    \alpha_{6l,\text{NS}} &= -\frac{6848}{63}, \\
    \alpha_{7l,\text{NS}} &= 0.0.
\end{flalign}

The spin parts of log-order-type PN coefficients
\begin{align}
&\alpha_{0l,\text{S}} = 0.0, \\
&\alpha_{1l,\text{S}} = 0.0, \\
&\alpha_{2l,\text{S}} = 0.0, \\
&\alpha_{3l,\text{S}} = 0.0, \\
&\alpha_{4l,\text{S}} = 0.0, \\
&\alpha_{5l,\text{S}} = \frac{1}{4536}\Big(-732985(\chi_{1L} + \chi_{2L} + \chi_{1L}\delta - \chi_{2L}\delta) \nonumber\\
&- 560(-1213(\chi_{1L} + \chi_{2L}) + 63(\chi_{1L} - \chi_{2L})\delta)\eta \nonumber\\
&+ 85680(\chi_{1L} + \chi_{2L})\eta^2\Big) \\
&\alpha_{6l,\text{S}} = 0.0, \\
&\alpha_{7l,\text{S}} = 0.0.
\end{align}

Equivalently, we have vanishing PN coefficients $\psi_1=\psi_{0l}=\psi_{1l}=\psi_{2l}=\psi_{3l}=\psi_{4l}=\psi_{7l}=0$. For the spin contribution part we have $\psi_{0,\text{S}}=\psi_{1,\text{S}}=\psi_{2,\text{S}}=\psi_{0l,\text{S}}=\psi_{1l,\text{S}}=\psi_{2l,\text{S}}=\psi_{3l,\text{S}}=\psi_{4l,\text{S}}=\psi_{6l,\text{S}}=\psi_{7l,\text{S}}=0$. So nontrivial PN coefficients that depend only on mass include $\psi_{0}$, $\psi_{2}$ and $\psi_{6l}$.

\end{appendix}

\bibliographystyle{unsrt}
\bibliography{refs}

\end{document}